\documentstyle[preprint,tighten,version2,aps]{revtex}

{\makeatletter%
\gdef\labeleqs#1{{%
\edef\@currentlabel{%
\ifappendixon\appletter\fi
\ifsecnumbers\ifnum\c@secnum>0
\arabic{secnum}.\fi\fi\arabic{equation}}%
\label{#1}%
}}%
}
\begin{document}
\draft
\preprint{IFUP-TH 29/93}
\begin{title}
Two dimensional $SU(N)\times SU(N)$ chiral models on the lattice
\end{title}
\author{Paolo Rossi and Ettore Vicari}
\begin{instit}
Dipartimento di Fisica dell'Universit\`a and
I.N.F.N., I-56126 Pisa, Italy
\end{instit}
\begin{abstract}
Lattice $SU(N)\times SU(N)$ chiral models are analyzed by strong and weak
coupling expansions and by numerical simulations.

$12^{th}$ order strong coupling series for the free and internal energy are
obtained for all $N\geq 6$. Three loop contributions to the internal energy
and to the lattice $\beta$-function are evaluated for all $N$ and
non-universal corrections to the asymptotic $\Lambda$ parameter are
computed in the ``temperature'' and the ``energy'' scheme.
Numerical simulations confirm a faster approach to asymptopia of
the energy scheme.
A phenomenological correlation between
the peak in the specific heat and the dip of the $\beta$-function
is observed.

Tests of scaling are performed for various physical quantities,
finding substantial scaling at $\xi \gtrsim 2$. In particular, at $N=6$
three different mass ratios are determined numerically and found in agreement,
within statistical errors of about 1\%,
with the theoretical predictions from the exact S-matrix theory.

\end{abstract}


\narrowtext
\section{Introduction}
\label{Introduction}
Two dimensional $SU(N)\times SU(N)$ principal chiral models
defined by the continuum lagrangian
\begin{equation}
L\;=\;{1\over T} \,{\rm Tr}\,\partial_\mu U \partial_\mu U^\dagger
\label{cont_action}
\end{equation}
are the simplest quantum field theories sharing with
non-abelian gauge theories the property of asymptotic freedom and whose
large N limit is a sum over planar diagrams.
Due to the existence of higher-order conservation laws, multiparticle
amplitudes are factorized, and exact $S$ matrices have been
proposed \cite{Abdalla,Wiegmann1,Wiegmann2}.
The resulting bound state mass spectrum is represented by
\begin{equation}
M_r\;=\;M\,{\sin (r\pi/N) \over \sin (\pi/N)}\;\;\;,\;\;\;\;\;1\le r\le
N-1\;\;\;,
\label{massratios}
\end{equation}
and the bound state of $r$ particles transforms as the totally
antisymmetric  tensor of rank $r$.
The mass-$\Lambda$ parameter ratio has been computed, and the result
is \cite{Balog}
\begin{equation}
{M\over\Lambda_{\overline {MS}}}\;=\;\sqrt{{8\pi\over e}} \;
{\sin \pi/N\over \pi/N}\;\;\;.
\label{mass-lambda}
\end{equation}
A ``standard'' lattice version of principal chiral models is obtained by
choosing the action
\begin{equation}
S\;=\;- 2\beta N \sum_{n,\mu} {\rm Re}\;{\rm Tr}\,[U_n U_{n+\mu}^\dagger
]\;\;\;,
\;\;\;\;\;\beta\;=\;{1\over NT}\;\;\;,
\label{latt-act}
\end{equation}
whose properties have been investigated by several authors
\cite{Green,Brihaye-Rossi,Shig,Shig2,Kogut,Kogut2,Guha}
especially by strong-coupling and mean field methods.
Numerical simulations have also been performed (at $N=3$), most recently
by Dagotto-Kogut \cite{Dagotto} and Hasenbusch-Meyer \cite{Hasenbusch}.

As a preliminary step within a more general program whose ultimate goal is
performing the numerical $1/N$ expansion of matrix-valued field theories,
we decided to explore the properties of principal chiral models at
larger-than-usual values of $N$.
In particular we wanted to investigate the following issues:

- region of applicability, accuracy and $N$ dependence of the strong
coupling series.

- onset of scaling, with special attention to the interplay between
thermodynamical (peak in the specific heat) and field theoretical (dip in
the $\beta$-function) effects.

- check of conjectured exact results (especially mass ratios) by
Monte Carlo measurements in the scaling region.

- role of coupling redefinitions in the widening of the asymptotic scaling
regions.

To this purpose, we performed a variety of strong coupling and
weak coupling calculations, and a number of numerical simulations for
different values of $N$, and especially at $N=6$, where the mass spectrum
is sufficiently non-trivial (two independent mass ratios can be measured
and compared with prediction), and $O(1/N^2)$ effects should be already
significantly depressed.

In the present paper we only report on our analytical
results, without offering any details on the derivations, that will be
presented
elsewhere.

\section{Analytical results}
\label{analyticalresults}

\subsection{Strong coupling}
\label{strongcoupling}

We found that the most convenient approach to strong coupling is the
character expansion. Free energy character expansion for $U(N)$ chiral
models to twelfth order and mass gap expansion to fifth order were presented
in Ref. \cite{Green}.
The formal extension of these series
to $SU(N)$ is easily achieved with the abovementioned precision
for $N > 6$. Paying
some attention in order to avoid double-counting, $SU(6)$  can be also
obtained by the same technique.

We found explicit representations of the coefficients of the $SU(N)$
character expansion in the strong coupling regime in terms of Bessel
functions, by generalizing the technique discussed in Refs.
\cite{Guha,Rossi}.
These representations are exact up to $O\left( \beta^{2N}\right)$.
As consequence we could compute the $SU(N)$ free energy to twelfth order
in $\beta$ for $N > 6$ in two dimensions
\begin{eqnarray}
&&{1\over N^2} F = 2\beta^2 \, + \, 2\beta^4 \, + \, 4\beta^6
\,+\, \left[ 14+{N^2(5N^2-2)\over (N^2-1)^2 }\right] \beta^8 \, + \, \left[56
+{8N^2 (5N^2-2)\over (N^2-1)^2 }\right]\beta^{10} \nonumber \\
&&+ \left[ 248+
{8N^2(35N^2-17)\over (N^2-1)^2 } + {2N^2(14N^6-11N^4+8N^2-2)\over (N^2-1)^4}
+{16N^4(9N^4-26N^2+8)\over 3(N^2-1)^2 (N^2-4)^2}\right] \beta^{12}
\nonumber \\
&&+\; O\left( \beta^{14}\right)\;\;+\;\;\;
4{N^{N-2}\over N!}\beta^N \,+\, \left[ 8 {N^{N-1}\over N!}-
4{N^N\over (N+1)!}\right]\beta^{N+2} \nonumber \\
&&+\;\left[ 2{N^{N+2}\over(N+2)!}
+ 4{N^{N+1}\over (N-1)N!} - 8{(N+2)N^N\over (N+1)!} +
24{N^{N-1}\over N!} + 4{N^{N-2}\over (N-2)!}\right]\beta^{N+4} \, + \,
O\left( \beta^{N+6}\right)\;.
\label{SCS1}
\end{eqnarray}
In the case $N=6$ an analysis of the $O\left( \beta^{N+6} \right)$
and $O\left( \beta^{2N}\right)$ contributions led to the result
\begin{equation}
{1\over N^2}F\;=\; 2\,\beta^2 \,+\, 2\,\beta^4 \,+\, 11.2\,\beta^6\,+\,
68.602449 \,\beta^8 \,+\, 374.945306\,\beta^{10} \,+\,
6395.760105 \,\beta^{12}\,+\,...
\label{SCS2}
\end{equation}
The internal energy (per link) density $E$ is immediately obtained from
the previous results by
\begin{equation}
E\;=\;1\,-\, {1\over 4N^2} {\partial F\over \partial \beta}\;\;\;.
\label{energy}
\end{equation}
These results have been used to draw the strong coupling curves in our
figures and compare very well with numerical simulations in the region
$\beta \lesssim 0.25 $.

\subsection{Weak coupling}
\label{weakcoupling}

Short weak coupling series for the free-energy density of $U(N)$ and $SU(N)$
chiral
models were presented in Ref. \cite{Brihaye-Rossi}.

We calculated the energy density up to three loops finding
\begin{equation}
E\;=\; 1\;-\;\langle \;{1\over N}
{\rm Re} \;{\rm Tr} \,[ U_n U^\dagger_{n+\mu} ]\;
\rangle \;=\; {N^2-1\over 8N^2\beta}\left[ 1\,+\,{a_1\over\beta}\,+\,
{a_2\over \beta^2}\,+\,...\;\right]
\label{WCenergy}
\end{equation}
where
\begin{eqnarray}
a_1&=& {N^2-2\over 32N^2} \;\;\;,\nonumber \\
a_2&=& {3N^4-14N^2+20\over 768N^4}\,+\, {N^4-4N^2+12\over 64N^4}Q_1
\,+\, {N^4-8N^2+24\over 64N^4}Q_2\;\;\;,
\end{eqnarray}
$Q_1$ and $Q_2$ being numerical constants: $Q_1=0.0958876$ and $Q_2=-0.0670$.

Asymptotic scaling requires the ratio of any dimensional quantity to the
appropriate power of the two loop lattice scale
\begin{equation}
\Lambda_{L,2l} \; = \; \left( b_0 T \right) ^{-b_1/b_0^2} \exp \left(
-{1\over b_0T} \right)
\label{ASYSC}
\end{equation}
to go to a constant as $T\rightarrow 0$.
$b_0$ and $b_1$ are the first universal coefficients of the expansion of the
$\beta$-function:
\begin{equation}
\beta_L(T) \;\equiv\; -a {d\over
da}T\;=\;-\,b_0\,T^2\,-\,b_1\,T^3\,-\,b_{2_L}\,T^4\,+\,...
\end{equation}
\begin{equation}
b_0\;=\;{N\over 8\pi}\;\;\;,\;\;\;\; b_1\;=\; {N^2\over 128\pi^2}\;\;\;.
\end{equation}
Evaluation of the ratios of $\Lambda$ parameters requires a one loop
calculation in perturbation theory, which leads to \cite{Shig2}:
\begin{equation}
{\Lambda_{\overline {MS}}\over \Lambda_L}\;=\; \sqrt{32} \exp \left(
\pi {N^2-2\over 2N^2} \right)\;\;\;.
\label{Lratio1}
\end{equation}

In order to get a more accurate description of the approach
to asymptotic scaling we performed the change of variables suggested by
Parisi \cite{Parisi}, defining  a new temperature $T_E$
proportional to the energy:
\begin{equation}
T_E \;=\; {8N\over N^2-1} \,E\;\;\;,\;\;\;\;\;\beta_E\;=\;{1\over
NT_E}\;\;\;.
\label{betae}
\end{equation}
Notice that the corresponding specific heat is, by definition,  constant.
The ratio of $\Lambda_E$, the $\Lambda$ parameter of the $\beta_E$
scheme, and $\Lambda_L$ is easily obtained from the two loop term
of the energy density:
\begin{equation}
{\Lambda_E\over\Lambda_L}\;=\;\exp \left(\pi {N^2-2\over 4N^2} \right)\;\;\;.
\label{ratioL2}
\end{equation}

We encountered the usual (and yet unexplained) phenomenon
of a much better convergence to asymptotic scaling for quantities plotted as
functions of $\beta_E$ \cite{Karsch,Wolff,amsterdam}.
We tried to check for a perturbative explanation of this phenomenon by
computing
the first perturbative correction to the two loop lattice scale
\begin{equation}
\Lambda\;=\; \left( 8\pi\beta\right)^{1/2}\,e^{-8\pi\beta}
\left[ 1+
{b_1^2-b_0b_2\over Nb_0^3}\beta^{-1} + O\left( \beta^{-2}\right) \right]\;\;\;,
\label{asy_corr}
\end{equation}
in the standard and the $\beta_E$ scheme, which requires the calculation
of the three-loop term of the $\beta$-function in both schemes.

In the standard scheme we found
\begin{equation}
b_{2_L}\;=\; {1\over (2\pi)^3} \,{N^3\over 128}\left[
1+\pi{N^2-2\over 2N^2}-\pi^2\left( {2N^4-13N^2+18\over 6N^4} +
4G_1\right)\right]
\label{b2L}
\end{equation}
where $G_1=0.04616363$ \cite{Falcioni,Campo-Rossi}.
The equivalence of the $SU(2)\times SU(2)$ chiral model to the $O(4)$
$\sigma$ model allows a check of this equation, indeed for $N=2$ it must give
(and indeed it does) the same $b_{2_L}$ of the standard lattice $O(4)$
$\sigma$ model \cite{Falcioni}.

The $\beta$-function of the $\beta_E$ scheme can be written
in the form
\begin{equation}
\beta_E(T_E)\;\equiv\;- a{d\over da}T_E\;=\; {8N^2\over N^2-1} \,C(T)
\,\beta_L(T)
\;\;\;,
\label{betaE}
\end{equation}
where
\begin{equation}
C(T)\;\equiv\;{1\over N} {dE\over dT}
\label{spec_heat}
\end{equation}
is the specific heat and $T$ must be considered as a function of $T_E$.
Expanding perturbatively Eq.~(\ref{betaE}) and using Eq.~(\ref{WCenergy})
one finds
\begin{equation}
b_{2_E}\;=\;b_{2_L} \,+\, N^2b_0\left( a_2-a_1^2\right) \,+\,
Nb_1a_1\;\;\;.
\label{b2E}
\end{equation}

As one may easily verify, the linear corrections to the two loop lattice
scale in Eq.~(\ref{asy_corr}) are small and of the same order of magnitude
(although of opposite sign). They cannot therefore explain the failure of
the first and the success of the second scheme with respect to achieving
asymptotic scaling. We believe that the origin of this phenomenon is fully
non-perturbative, and it can presumably be traced to the phenomenologically
apparent correlation existing between the peak in the specific heat and the
dip in the lattice $\beta$-function: the non-perturbative variable
transformation that flattens the peak manages to fill
the dip, in a theoretically yet uncontrolled way.

\section{Numerical results}
\label{numericalresults}

We performed Monte Carlo simulations of the lattice $SU(N)\times SU(N)$
chiral models for a wide range of values of $N$ (in particular
$N=3,6,9,15$) and $\beta$. Summaries of the runs are presented in Tables
\ref{N3-table},\ref{N6-table},\ref{N9-table} and \ref{N15-table}.

In our simulations we implemented the Cabibbo-Marinari algorithm
\cite{Cabibbo} to upgrade $SU(N)$ matrices by updating its $SU(2)$
subgroups.
In most cases, we chose to update
the $N-1$ diagonal subsequent $SU(2)$ subgroups of each $SU(N)$ matrix
variable by employing
the over-heat-bath algorithm \cite{Petronzio}
(for the ``heat bath'' part of it we used the Kennedy-Pendleton
algorithm \cite{Kennedy}).

An important class of observables of the $SU(N)\times SU(N)$ chiral models
can be constructed by considering the group invariant correlation function
\begin{equation}
G(x-y)\;=\;\langle \;{1\over N} {\rm Re} \;{\rm Tr} \,[ U(x) U(y)^\dagger
]\;\rangle\;\;\;.
\end{equation}
We define the correlation function $\xi_G$ from the second moment of the
correlation function $G(x)$. On the lattice
\begin{equation}
\xi_G^2 = {1\over4\sin^2\pi/L} \,
\left[{\widetilde G(0,0)\over\widetilde G(0,1)} - 1\right]\;\;\;,
\label{xiG}
\end{equation}
where $\widetilde G(k_x,k_y)$ is the Fourier transform of $G(x)$.
The inverse mass gap $\xi_w$ is extracted from the long distance behavior
of the zero space momentum correlation function constructed with $G(x)$.
Moreover we measured the diagonal wall-wall correlation length
$\xi_d$ to test rotation invariance.
$M\equiv 1/\xi_w$ should reproduce in the continuum limit the mass of
the fundamental state.
The first definition of correlation length $\xi_G$ offers the advantage of
being directly measurable, while the calculation of $\xi_w$ requires
a fit procedure. On the other hand, since $\xi_G$ is an off-shell quantity
an analytical prediction exists only for
the inverse mass-gap (Eq.~(\ref{mass-lambda})).

In Tables \ref{N3-table},\ref{N6-table},\ref{N9-table} and \ref{N15-table}
 we present data for the energy density $E$,
the specific heat $C\equiv {1\over N} {dE\over dT}$,
the magnetic susceptibility $\chi_m$
defined from the correlation function $G(x)$,
the correlation length $\xi_G$, the dimensionless ratios $\xi_G/\xi_w$
and $\xi_d/\xi_w$,
respectively for $N=3,6,9,15$.

We carefully checked for finite size  effects.
It turned out that for $z\equiv L/\xi_G\gtrsim  8$ the finite size
systematic errors
in evaluating infinite volume quantities should be safely smaller than 1\%,
which is the typical statistical error of our data.

In Figs.~\ref{en_N6} and \ref{en_N9}
we show the energy density versus $\beta$ respectively at
$N=6$ and $N=9$. There the strong coupling series up to twelfth order
in $\beta$ and the weak coupling one up to third order in $\beta^{-1}$
are drawn.

As in other asymptotically free models, at all values of $N$
the specific heat shows a peak, connecting the two different asymptotic
behaviors:
monotonically increasing in the strong coupling region and
decreasing at large $\beta$.
In Figs.~\ref{asy_N3},\ref{asy_N6},\ref{asy_N9} and \ref{asy_N15}
$C$ is plotted respectively for $N=3,6,9,15$ with
the corresponding $13^{th}$ order strong coupling series (except for $N=3$).
Increasing $N$, the peak moves slightly towards higher
$\beta$ values ($\beta_{peak}\simeq 0.285$ at $N=6$, $\beta_{peak}\simeq
0.30$ at $N=15$), and becomes more and more pronounced.
We found the position of the peak to be more stable
at large $N$ when plotting $C$ versus $\xi_G$, as in Fig.~\ref{specheat}.
Notice that, increasing $N$,  the specific heat around the peak
does not show any apparent convergence to a finite value,
which might be an indication of a (first order?) phase transition at
$N=\infty$.

The $12^{th}$ order ($13^{th}$ order) strong coupling series of the energy
(specific heat) are in quantitative agreement (within our statistical
errors) for $\beta\lesssim 0.2$, and in qualitative agreement up to the peak
of the specific heat, whose position should give an estimate of
the strong coupling convergence radius.

Tests of scaling, based on the stability of dimensionless physical
quantities (for example, the ratio $\xi_G/\xi_w$) and rotation invariance
(checking that $\xi_w/\xi_d\simeq 1$), showed that, within our statistical
errors,
the scaling region is reached already at small correlation lengths,
i.e. for $\xi_G\simeq 2$.
Fitting data in the scaling region to a constant we found
\begin{eqnarray}
\xi_G/\xi_w &=& 0.987(2) \;\;\;{\rm for} \;\;\;N=3\;\;\;,\nonumber \\
&=& 0.993(2) \;\;\;{\rm for} \;\;\;N=6\;\;\;,\nonumber \\
&=& 0.995(3) \;\;\;{\rm for} \;\;\;N=9\;\;\;,\nonumber \\
&=& 0.994(4) \;\;\;{\rm for} \;\;\;N=15\;\;\;.
\label{xiratio}
\end{eqnarray}
Notice that scaling is observed even before the peak of the specific heat.
Since strong coupling series should be effective in this region,
it might be possible to calculate continuum physical
quantities by strong coupling techniques.
In order to investigate this issue work to extend the strong coupling series is
in progress.

We checked asymptotic scaling according to the two loop formula
(\ref{ASYSC}) by analyzing $M_G/\Lambda_{L,2l}\equiv 1/(\xi_G \Lambda_{L,2l})$.
In Figs.~\ref{asy_N3},\ref{asy_N6},\ref{asy_N9} and \ref{asy_N15}
we show the corresponding data respectively for $N=3,6,9,15$.
At all values of $N$ we observe the usual dip in the $\beta$-function,
which is, again, more and more pronounced when increasing $N$. Since
$\xi_G\simeq\xi_w$ we compare $M_G/\Lambda_{L,2l}$ directly with
Eq.~(\ref{mass-lambda}) (using also Eq.~(\ref{Lratio1})),
whose predictions,
\begin{eqnarray}
{M/\Lambda_L}&=&  48.266... \;\;\;{\rm for} \;\;\;N=3\;\;\;,\nonumber \\
&=& 72.412... \;\;\;{\rm for} \;\;\;N=6\;\;\;,\nonumber \\
&=& 77.989...\;\;\;{\rm for} \;\;\;N=9\;\;\;,\nonumber \\
&=& 81.001... \;\;\;{\rm for} \;\;\;N=15\;\;\;,
\end{eqnarray}
are represented by dashed lines in the figures.
Notice that the Monte Carlo data are much larger than
the predicted values, while the first perturbative corrections
in Eq.~(\ref{asy_corr}) are, in all cases, about 20\% at $\beta\simeq 0.3$.
Furthermore, data show a similarity with the behavior of the specific
heat, strengthening the idea of a strong correlation between the two
phenomena.

The approach to asymptotic scaling gets an impressive
improvement using the $\beta_E$ scheme.
In Figs.~\ref{asy_N3},\ref{asy_N6},\ref{asy_N9} and \ref{asy_N15}
we also plot
\begin{equation}
{M_G/\Lambda_{L,2l}}|_E\;\equiv\; {1\over \xi_G \Lambda_{E,2l}}\times
{\Lambda_E\over \Lambda_L}\;\;\;.
\label{MGE}
\end{equation}
Now data approach the correct value, and the discrepancies are even smaller
than the linear correction calculated in Sec. \ref{numericalresults}
(which is about 15\% at $\beta\simeq 0.3$).
So flattening the peak of the specific heat by performing the coupling
redefinitions $T\rightarrow T_E$, the dip of the $\beta$-function
disappears.
We believe this to be the key point of the success of the
$\beta_E$ scheme in widening the asymptotic scaling region.
The peak of the specific heat should be explicable
in terms of complex $\beta$-singularities of the partition function
close to the real axis \cite{Marinari}.
The sharpening of the peak with increasing $N$ would indicate that the complex
singularities get nearer and nearer to the real axis, pinching it at
$N=\infty$ where a phase transition is expected.
Such singularities should also cause the abrupt departure from the weak
coupling behavior.
Then a coupling transformation
eliminating the peak should move the complex $\beta$-singularities
away from the real axis, and therefore improve the approach to asymptotic
scaling.

{}From the Monte Carlo data and the exact result (\ref{mass-lambda})
we can extract the effective $\Lambda$-parameters $\Lambda_L(N,\beta)$
and $\Lambda_E(N,\beta_E)$. Fig.~\ref{LL} and \ref{LE}
show respectively the ratios
$\Lambda_L(N,\beta)/\Lambda_{L,2l}(N,\beta)$ and
$\Lambda_E(N,\beta_E)/\Lambda_{E,2l}(N,\beta_E)$, where
$\Lambda_{L,2l}(N,\beta)$ and $\Lambda_{E,2l}(N,\beta_E)$ are the
corresponding two loop functions:
$\Lambda_{L,2l}(N,x)=\Lambda_{E,2l}(N,x)=(8\pi x)^{1/2}\exp (-8\pi x)$.
Similarly to the specific heat,
the effective $\Lambda$-parameter
$\Lambda_L(N,\beta)$ does not give evidence of convergence at large $N$.
On the contrary $\Lambda_E(N,\beta_E)$ appears to approach a finite function
$\Lambda_E(\infty,\beta_E)$, which is well approximated by the two
loop formula.

In conclusion, scaling and asymptotic scaling (in the $\beta_E$ scheme)
are observed at all values of $N$ considered, even around the peak of the
specific heat. It is interesting to notice that,
even though the behavior of the specific heat with respect to $N$
suggests the existence of a phase transition at $N=\infty$,
the above scenario is apparently stable at large $N$.

\section{Mass spectrum at $N=6$}
\label{massspectrum}

We studied the mass spectrum at $N=6$, where Eq.~(\ref{massratios})
predicts the existence of two independent mass ratios.
In order to extract the other two independent mass values besides the
fundamental one, we considered the following operators:
\begin{equation}
O^{(2)}_{abcd}\;=\;U_{ab}U_{cd}-U_{ad}U_{cb}\;\;\;,
\label{O2}
\end{equation}
\begin{equation}
O^{(3)}_{abcdef}\,=\,U_{ab}U_{cd}U_{ef}
-U_{ab}U_{cf}U_{ed}-U_{ad}U_{cb}U_{ef}
+ U_{ad}U_{cf}U_{eb}+U_{af}U_{cb}U_{ed}-U_{af}U_{cd}U_{eb}\;\;,
\label{O3}
\end{equation}
having respectively the same transformation properties of the two
and three particle bound states.
The mass values $M_2$ and $M_3$
were determined from the large distance behavior of the
zero space momentum correlation functions constructed with the above operators.
In practice we found distances $d\gtrsim 1.5\,\xi_G$
to be large enough to fit the data to the expected exponential behavior.
In Table \ref{mass-table} and in Fig. \ref{mass-fig}
we present the data for the ratios
$M/M_G$, $M_2/M$ and $M_3/M$, analyzed using the jackknife method.
They show good scaling. Fitting them to a constant we found
\begin{eqnarray}
{M/M_G} &=& 0.993(2) \;\;\;, \nonumber \\
{M_2/ M} &=& 1.74(1) \;\;\;, \nonumber \\
{M_3/ M} &=& 2.01(2)\;\;\;.
\label{massres}
\end{eqnarray}
This result confirms, within statistical errors of about 1\%, the conjectured
exact
result (\ref{massratios}), which predicts
\begin{eqnarray}
{M_2/ M} &=& 1.73205... \;\;\;, \nonumber \\
{M_3/ M} &=& 2\;\;\;.
\label{masspred}
\end{eqnarray}




\figure{Energy versus $\beta$ for $N=6$.
The dashed and dotted lines
represent respectively the $12^{th}$ order strong coupling
and the third order weak coupling series.
\label{en_N6}}

\figure{Energy versus $\beta$ for $N=9$.
The dashed and dotted lines
represent respectively the $12^{th}$ order strong coupling
and the third order weak coupling series.
\label{en_N9}}

\figure{Specific heat and asymptotic scaling test $M_G/\Lambda_{L,2l}$ for N=3.
The dashed line shows the analytical prediction (\ref{mass-lambda}).
Data of $C$ are multiplied by 500.
\label{asy_N3}}

\figure{Specific heat and asymptotic scaling test $M_G/\Lambda_{L,2l}$ for N=6.
The dotted line represents the $13^{th}$ order strong coupling series of $C$.
The dashed line shows the analytical prediction (\ref{mass-lambda}).
Data of $C$ and the corresponding strong coupling series are multiplied by 500.
\label{asy_N6}}

\figure{Specific heat and asymptotic scaling test $M_G/\Lambda_{L,2l}$ for N=9.
The dotted line represents the $13^{th}$ order strong coupling series of $C$.
The dashed line shows the analytical prediction (\ref{mass-lambda}).
\label{asy_N9}}

\figure{Specific heat and asymptotic scaling test $M_G/\Lambda_{L,2l}$ for
N=15.
The dotted line represents the $13^{th}$ order strong coupling series of $C$.
The dashed line shows the analytical prediction (\ref{mass-lambda}).
\label{asy_N15}}

\figure{Specific heat versus $\xi_G$.
\label{specheat}}

\figure{$\Lambda_L(N,\beta)/\Lambda_{L,2l}(N,\beta)$ versus $\beta$.
\label{LL}}

\figure{$\Lambda_E(N,\beta_E)/\Lambda_{E,2l}(N,\beta_E)$ versus $\beta_E$.
\label{LE}}

\figure{ $M/M_G$, $M_2/M$ and $M_3/M$ versus $\xi_G$ for $N=6$.
Dotted lines show the exact predictions (\ref{massratios}) for
the ratios $M_2/M$ and $M_3/M$.
\label{mass-fig}}


\begin{table}
\caption{Summary of the numerical results for N=3.}
\label{N3-table}
\begin{tabular}{r@{}lrr@{}lr@{}lr@{}lr@{}lr@{}lr@{}lr@{}l}
\multicolumn{2}{c}{$\beta$}&
\multicolumn{1}{r}{$L$}&
\multicolumn{2}{c}{$E$}&
\multicolumn{2}{c}{$C$}&
\multicolumn{2}{c}{$\chi_m$}&
\multicolumn{2}{c}{$\xi_G$}&
\multicolumn{2}{c}{$\xi_G/\xi_w$} &
\multicolumn{2}{c}{$\xi_d/\xi_w$} \\
\tableline
0&.18 & 18  & 0&.74118(5) & 0&.0712(4)   & 3&.843(5) & 1&.003(17) && &&\\
0&.225& 24  & 0&.62819(12)& 0&.143(2)    & 9&.15(3)  & 1&.87(2)   & 0&.990(10)&
0&.994(13) \\
0&.25 & 30  & 0&.55589(4) & 0&.1814(9)   & 19&.09(4) & 3&.027(14) & 0&.986(5) &
0&.995(7) \\
0&.27 & 36  & 0&.49992(4) & 0&.1935(10)  & 40&.22(9) & 4&.79(2)   & 0&.988(3) &
1&.004(4)  \\
0&.27 & 42  & 0&.50000(3) & 0&.1936(10)  & 40&.05(9) & 4&.78(2)   & 0&.987(5) &
0&.999(3)  \\
0&.27 & 48  & 0&.50003(3) & 0&.1918(12)  & 40&.21(9) & 4&.81(3)   & 0&.984(6) &
0&.995(6) \\
0&.29 & 81  & 0&.45172(3) & 0&.187(2)    & 93&.1(5)  & 7&.99(10)  & 0&.989(10)&
1&.000(10)  \\
0&.30 & 90  & 0&.43111(2) & 0&.176(2)    &144&.2(8)  & 10&.41(10) & 0&.993(7) &
0&.989(9)  \\
0&.315&120  & 0&.40400(2) & 0&.166(3)    &283&(3)    & 15&.5(3)   & 0&.981(9) &
0&.997(12) \\
\end{tabular}
\end{table}

\begin{table}
\caption{Summary of the numerical results for N=6.}
\label{N6-table}
\begin{tabular}{r@{}lrr@{}lr@{}lr@{}lr@{}lr@{}lr@{}lr@{}l}
\multicolumn{2}{c}{$\beta$}&
\multicolumn{1}{r}{$L$}&
\multicolumn{2}{c}{$E$}&
\multicolumn{2}{c}{$C$}&
\multicolumn{2}{c}{$\chi_m$}&
\multicolumn{2}{c}{$\xi_G$}&
\multicolumn{2}{c}{$\xi_G/\xi_w$} &
\multicolumn{2}{c}{$\xi_d/\xi_w$} \\
\tableline
0&.10  & 15  & 0&.89781(3) & 0&.01076(7) & 1&.5717(12) & 0&.396(12) &&  &&\\
0&.15  & 18  & 0&.84170(2) & 0&.02690(10)& 2&.091(2)   & 0&.56(2)   &&  &&\\
0&.175 & 18  & 0&.81060(2) & 0&.0399(2)  & 2&.4806(14) & 0&.671(9)  &&  && \\
0&.20  & 24  & 0&.77592(2) & 0&.0596(3)  & 3&.0425(18) & 0&.816(12) &&  &&  \\
0&.225 & 18  & 0&.73506(4) & 0&.0903(6)  & 3&.936(3)   && &&  && \\
0&.25  & 18  & 0&.68234(8) & 0&.153(2)   & 5&.731(9)   & 1&.332(10) &
0&.987(11) & 0&.998(15) \\
0&.26  & 24  & 0&.65537(7) & 0&.200(3)   & 7&.120(11)  & 1&.560(14) && &&  \\
0&.27  & 24  & 0&.62377(5) & 0&.248(2)   & 9&.407(9)   & 1&.882(6)  & 0&.991(7)
& 0&.998(7) \\
0&.28  & 30  & 0&.58690(8) & 0&.302(4)   &13&.68(3)    & 2&.410(17) & 0&.994(6)
& 0&.997(6)  \\
0&.285 & 30  & 0&.56710(7) & 0&.332(6)   &17&.28(3)    & 2&.836(13) & 0&.989(6)
& 1&.007(7) \\
0&.29  & 30  & 0&.54730(7) & 0&.323(4)   &22&.31(5)    & 3&.354(18) & 0&.991(8)
& 0&.992(5)\\
0&.29  & 36  & 0&.54732(7) & 0&.323(5)   &22&.38(6)    & 3&.37(2)   & 0&.986(6)
& 0&.996(8)\\
0&.30  & 36  & 0&.51134(5) & 0&.300(4)   &38&.27(9)    & 4&.75(2)   & 0&.989(4)
& 0&.996(4) \\
0&.30  & 42  & 0&.51134(5) & 0&.296(3)   &38&.30(12)   & 4&.76(2)   & 0&.994(6)
& 0&.998(10) \\
0&.30  & 48  & 0&.51139(8) & 0&.288(8)   &38&.23(16)   & 4&.78(5)   & 0&.999(6)
& 0&.995(5) \\
0&.31  & 54  & 0&.48188(4) & 0&.256(3)   &65&.1(3)     & 6&.55(5)   & 0&.986(6)
& 1&.005(5) \\
0&.31  & 60  & 0&.48186(4) & 0&.258(4)   &65&.1(2)     & 6&.61(4)   & 0&.992(5)
& 1&.000(7) \\
0&.32  & 75  & 0&.45764(3) & 0&.227(3)   &108&.4(5)    & 9&.14(7)   & 0&.996(5)
& 0&.995(7) \\
0&.32  & 81  & 0&.45769(3) & 0&.230(5)   &107&.4(6)    & 9&.00(11)  & 0&.997(7)
& 1&.001(8) \\
0&.40  & 60  & 0&.34033(5) & &   & &    & &  & &  && \\
0&.50  & 60  & 0&.26335(4) & &   & &    & &  & &  && \\
\end{tabular}
\end{table}

\begin{table}
\caption{Summary of the numerical results for N=9.}
\label{N9-table}
\begin{tabular}{r@{}lrr@{}lr@{}lr@{}lr@{}lr@{}lr@{}lr@{}l}
\multicolumn{2}{c}{$\beta$}&
\multicolumn{1}{r}{$L$}&
\multicolumn{2}{c}{$E$}&
\multicolumn{2}{c}{$C$}&
\multicolumn{2}{c}{$\chi_m$}&
\multicolumn{2}{c}{$\xi_G$}&
\multicolumn{2}{c}{$\xi_G/\xi_w$} &
\multicolumn{2}{c}{$\xi_d/\xi_w$} \\
\tableline
0&.175  & 18  & 0&.812908(11) & 0&.03769(13) & 2&.4494(12) & 0&.672(7) &&  &&\\
0&.20   & 24  & 0&.780870(12) & 0&.0538(3)   & 2&.948(12)  & 0&.792(10)&&  &&\\
0&.25   & 18  & 0&.70339(8)   & 0&.120(2)    & 4&.868(8)   & 1&.177(11)&&  &&\\
0&.27   & 24  & 0&.65937(13)  & 0&.205(6)    & 6&.792(15)  & 1&.497(17)&
0&.980(8)  & 0&.995(10)\\
0&.28   & 24  & 0&.62920(12)  & 0&.268(8)    & 8&.78(2)    & 1&.777(13)&
0&.990(9)  & 0&.997(10)\\
0&.29   & 30  & 0&.58801(13)  & 0&.410(15)   &13&.27(3)    & 2&.371(14)&
0&.994(7)  & 0&.987(7) \\
0&.295  & 36  & 0&.56281(12)  & 0&.457(16)   &17&.96(5)    & 2&.89(2)  &
0&.994(7)  & 0&.998(5) \\
0&.30   & 30  & 0&.53844(21)  & 0&.420(25)   &25&.36(14)   & 3&.67(3)  & &
&&\\
0&.30   & 36  & 0&.53845(16)  & 0&.411(21)   &25&.22(12)   & 3&.63(4)  &
0&.992(8) & 0&.998(8)\\
0&.30   & 42  & 0&.53847(12)  & 0&.411(16)   &25&.26(9)    & 3&.68(4)  &
0&.999(6) & 1&.009(6) \\
0&.31   & 42  & 0&.50028(9)   & 0&.307(12)   &47&.1(2)     & 5&.39(5)  && &&\\
0&.31   & 54  & 0&.50035(9)   & 0&.318(12)   &47&.2(2)     & 5&.46(5)  &
0&.993(9) & 0&.995(10) \\
0&.32   & 60  & 0&.47234(5)   & 0&.252(10)   &81&.6(4)     & 7&.65(7)  &
0&.995(4) & 1&.006(5) \\
\end{tabular}
\end{table}

\begin{table}
\caption{Summary of the numerical results for N=15.}
\label{N15-table}
\begin{tabular}{r@{}lrr@{}lr@{}lr@{}lr@{}lr@{}lr@{}lr@{}l}
\multicolumn{2}{c}{$\beta$}&
\multicolumn{1}{r}{$L$}&
\multicolumn{2}{c}{$E$}&
\multicolumn{2}{c}{$C$}&
\multicolumn{2}{c}{$\chi_m$}&
\multicolumn{2}{c}{$\xi_G$}&
\multicolumn{2}{c}{$\xi_G/\xi_w$} &
\multicolumn{2}{c}{$\xi_d/\xi_w$} \\
\tableline
0&.20  & 24  & 0&.781395(16)& 0&.0515(5) & 2&.9392(14)& 0&.775(12)&  & &&\\
0&.25  & 24  & 0&.70846(3)  & 0&.100(3)  & 4&.692(4)  & 1&.158(14)&  & &&\\
0&.28  & 24  & 0&.64990(9)  & 0&.204(8)  & 7&.251(11) & 1&.561(10)& 0&.997(10)&
0&.992(9)\\
0&.29  & 24  & 0&.62134(13) & 0&.294(14) & 9&.31(2)   & 1&.857(9) & 0&.998(6) &
0&.995(7)\\
0&.30  & 30  & 0&.56809(18) & 0&.67(4)   &16&.53(5)   & 2&.750(12)& 0&.995(5) &
1&.000(6)\\
0&.31  & 45  & 0&.51202(10) & 0&.35(3)   &38&.89(16)  & 4&.78(3)  & 0&.983(8) &
0&.998(7)\\
\end{tabular}
\end{table}

\begin{table}
\caption{ Mass spectrum for N=6.}
\label{mass-table}
\begin{tabular}{r@{}lr@{}lr@{}lr@{}l}
\multicolumn{2}{c}{$\beta$}&
\multicolumn{2}{c}{$M/M_G$}&
\multicolumn{2}{c}{$M_2/M$}&
\multicolumn{2}{c}{$M_3/M$} \\
\tableline
0&.29 & 0&.991(8) & 1&.78(3)  & 2&.02(3) \\
0&.30 & 0&.993(3) & 1&.74(2)  & 2&.04(4) \\
0&.31 & 0&.992(4) & 1&.72(2)  & 1&.98(3) \\
0&.32 & 0&.996(4) & 1&.74(2)  & 2&.06(5) \\
\end{tabular}
\end{table}


\begin{references}

\bibitem{Abdalla} E.~Abdalla, M.C.B.~Abdalla and A.~Lima-Santos,
Phys.\ Lett.\ {\bf 140B}, 71 (1984).

\bibitem{Wiegmann1} P.~Wiegmann,
Phys.\ Lett.\ {\bf 141B}, 217 (1984).


\bibitem{Wiegmann2} P.~Wiegmann,
Phys.\ Lett.\ {\bf 142B}, 173 (1984).

\bibitem{Balog}J.~Balog, S.~Naik, F.~Niedermayer, and
P.~Weisz,
Phys.\ Rev.\ Lett.\ {\bf 69}, 873 (1992).

\bibitem{Green} F.~Green and S. Samuel, Nucl.\ Phys.\
{\bf B190}, 113 (1981).

\bibitem{Brihaye-Rossi} Y.~Brihaye and P.~Rossi,
Nucl.\ Phys.\ {\bf B235}, 226 (1984).

\bibitem{Shig} J.~Shigemitsu, J.B.~Kogut and D.K.~Sinclair,
Phys.\ Lett.\ {\bf 100B}, 316 (1981).

\bibitem{Shig2} J.~Shigemitsu and J.B.~Kogut,
Nucl.\ Phys.\ {\bf B190}, 365 (1981).

\bibitem{Kogut} J.~Kogut, M.~Snow and M.~Stone,
Phys.\ Rev.\ Lett.\ {\bf 47}, 1767 (1981).

\bibitem{Kogut2} J.~Kogut, M.~Snow and M.~Stone,
Nucl.\ Phys.\ {\bf B215}, 45 (1983).

\bibitem{Guha}A.~Guha and S.C.~Lee,
Nucl.\ Phys.\ {\bf B240}, 141 (1984).

\bibitem{Dagotto} F.~Dagotto and J.B.~Kogut,
Nucl.\ Phys.\ {\bf B290}, 451 (1987).

\bibitem{Hasenbusch}M.~Hasenbusch and S.~Meyer,
Phys.\ Rev.\ Lett.\ {\bf 68}, 435 (1992).

\bibitem{Rossi} P.~Rossi,
Phys.\ Lett.\ {\bf 117B}, 72 (1982).

\bibitem{Parisi}G.~Parisi, Proceedings of the XXth
Conf. on High energy physics (Madison, WI 1980).

\bibitem{Karsch} F.~Karsch and R.~Petronzio,
Phys.\ Lett.\ {\bf 139}, 403 (1984).

\bibitem{Wolff}U.~Wolff,
Phys.\ Lett.\ {\bf 248}, 335 (1990).

\bibitem{amsterdam} M.~Campostrini, P.~Rossi and E.~Vicari,
Phys. Rev. {\bf D 43} (1992) 4643.

\bibitem{Falcioni}M.~Falcioni and A.~Treves,
Nucl.\ Phys.\ {\bf B265}, 671 (1986).

\bibitem{Campo-Rossi} M.~Campostrini and P.~Rossi,
Phys.\ Lett.\ {\bf 242B}, 225 (1990).


\bibitem{Cabibbo}N.~Cabibbo and E.~Marinari,
Phys.\ Lett.\ {\bf 119B}, 387 (1982).

\bibitem{Petronzio}R.~Petronzio and E.~Vicari,
Phys.\ Lett.\  {\bf 254B}, 444 (1991).

\bibitem{Kennedy}A.D.~Kennedy and B.J.~Pendleton,
Phys.\ Lett.\ {\bf 156B}, 393 (1985).


\bibitem{Marinari}M.~Falcioni, E.~Marinari, M.L.~Paciello, G.~Parisi and
B.~Taglienti, Phys.\ Lett.\ {\bf 102B}, 270 (1981);
Phys.\ Lett.\ {\bf 108B}, 331 (1982).

\end{references}
\end{document}